\documentclass[twoside,final,1p,times]{elsarticle}
\usepackage{xcolor}
\usepackage{hyperref}
\usepackage{parskip}
\usepackage{booktabs}
\usepackage{longtable}
\usepackage{array}
\usepackage{colortbl}
\usepackage{tcolorbox}
\usepackage{graphicx}
\usepackage{amsmath}
\usepackage{amssymb}
\usepackage{caption}
\usepackage{subcaption}
\usepackage{listings}
\usepackage{enumitem}

\tcbuselibrary{breakable, skins, theorems}
\usetikzlibrary{positioning,shapes.misc}

\journal{Data Intelligence}
\hypersetup{pdftitle={Biomedical Knowledge Composition: A Software Engineering Perspective},
  pdfauthor={Natallia Kokash, Adam S.Z. Belloum, Paola Grosso}}

\definecolor{codered}{RGB}{160,20,30}  

\newcommand{\code}[1]{\texttt{\small\color{codered}#1}}
\newcommand{\term}[1]{\textit{#1}}

\usepackage[T1]{fontenc}
\usepackage{helvet}

\definecolor{rootblue}{HTML}{174E9C}
\definecolor{cOrange}{HTML}{C47A33}
\definecolor{cGreen}{HTML}{4E8A3E}
\definecolor{cPurple}{HTML}{6A5399}
\definecolor{cRed}{HTML}{C0504A}
\definecolor{conn}{HTML}{6F8AAE}
\definecolor{rule}{HTML}{DADADA}
\definecolor{txt}{HTML}{3C4A5C}

\tikzset{
  root/.style={rounded corners=5pt, fill=rootblue, text=white,
    font=\bfseries\large, align=center, inner sep=0pt,
    minimum width=4.7cm, minimum height=1.55cm},
  cat/.style={rounded corners=5pt, text=white, font=\bfseries\normalsize,
    align=center, inner sep=0pt, minimum height=1.2cm},
  line/.style={draw=conn, line width=0.9pt},
  item/.style={anchor=west, text=txt, font=\small, align=left, inner sep=0pt},
  sep/.style={draw=rule, line width=0.5pt},
}

\newcommand{\branch}[8]{%
  \draw[line] (#2,305) -- (#2,#5) -- ++(33,0);
  \draw[line] (#2,#3) -- ++(33,0);
  \draw[line] (#2,#4) -- ++(33,0);
  \foreach \yy/\lab in {#3/{#6}, #4/{#7}, #5/{#8}}{
    \fill[#1] (#2+32,\yy) circle (8);
    \node[item] at (#2+55,\yy) {\lab};
    \draw[sep] (#2+47,\yy+37) -- ++(237,0);
  }
}

\definecolor{kgblue}{RGB}{16,64,160}
\definecolor{kggreen}{RGB}{16,128,48}

\newcommand{\sect}[7]{%
  \begin{scope}[shift={(#1,#2)}]
    \draw[rounded corners=2pt, draw=#3!50, fill=white, line width=0.6pt] (0,0) rectangle (8.4,-2.2);
    \node[anchor=north west, inner sep=0pt, fill=#3, text=white, font=\bfseries\small,
          minimum width=8.4cm, minimum height=0.6cm] at (0,0) {#4};
    \node[item, anchor=north west] at (0.2,-0.85) {#5};
    \node[item, anchor=north west] at (3.0,-0.85) {#6};
    \node[item, anchor=north west] at (5.8,-0.85) {#7};
  \end{scope}%
}

\newcommand{\column}[3]{%
  \draw[rounded corners=3pt, draw=#2!50, fill=white, line width=0.8pt] (#1,0) rectangle (#1+9.2,-13.8);
  \node[anchor=north west, inner sep=0pt, fill=#2, text=white, font=\bfseries\normalsize,
        minimum width=9.2cm, minimum height=1.0cm] at (#1,0) {#3};
}

\newtcolorbox{gapbox}[2][]{
  enhanced, breakable,
  boxrule=0.5pt, arc=2pt,
  colback=white, colframe=black!40,
  title={\textbf{#2}},
  fonttitle=\small,
  left=8pt, right=8pt, top=6pt, bottom=6pt,
  before skip=10pt, after skip=8pt,
  #1
}

\newcounter{sechallenge}
\newcommand{\sechallenge}[1]{%
  \stepcounter{sechallenge}%
  \subsection{SE Challenge \thesechallenge: #1}%
}

\newcolumntype{L}[1]{>{\raggedright\arraybackslash}p{#1}}
\newcolumntype{C}[1]{>{\centering\arraybackslash}p{#1}}

\begin{document}

\begin{frontmatter}

\title{Biomedical Knowledge Composition: A Software Engineering Perspective}

\author[inst1]{Natallia Kokash\fnref{orcid1}}
\author[inst1]{Adam S.Z. Belloum\fnref{orcid2}}
\author[inst1]{Paola Grosso\fnref{orcid3}}

\address[inst1]{Informatics Institute, University of Amsterdam, The Netherlands}

\fntext[orcid1]{ORCID:\href{https://orcid.org/0000-0003-3639-1245}{0000-0003-3639-1245}}
\fntext[orcid2]{ORCID:\href{https://orcid.org/0000-0001-6306-6937}{0000-0001-6306-6937}}
\fntext[orcid3]{ORCID:\href{https://orcid.org/0000-0003-4600-9812}{0000-0003-4600-9812}}

\begin{abstract}
Biomedical research has accumulated large volumes of molecular, clinical, and population
data, yet translating this wealth into actionable knowledge remains severely constrained
by technical and organizational difficulties. This article presents a unified treatment
of two complementary perspectives on biomedical knowledge infrastructure.

The first perspective introduces the biomedical domain to software engineers: it explains
why knowledge graphs (KGs) are the central integrative data structure in modern biomedicine,
characterizes five core data harmonization challenges (identifier mapping, entity resolution,
schema alignment, evidence integration, and provenance tracking), surveys principal
application domains from drug discovery to digital twins, profiles six representative
KG systems with contrasting design choices, and reports a first-hand attempt to reuse a
published KG.

The second perspective asks why engineering biomedical knowledge infrastructure remains
so difficult. We argue that a contributing root cause is limited adoption of software
tooling and practices that make development in other mature domains, particularly web
engineering, reliably composable and reproducible: package management, typed namespaces,
canonical interchange formats, service composition protocols, reproducible pipelines, and
lifecycle governance. Against this backdrop, eight open engineering challenges for
biomedical data integration are cataloged, each with partial solutions but no universally
adopted stack yet assembled.

The article focuses on the \emph{reproducible process of assembling KGs}: reusable build
pipelines, versioned dependencies, and engineering practices that let others compile and
customize a KG from source. The two perspectives provide
domain grounding for newcomers and a research agenda for software engineers seeking to
contribute to biomedical knowledge infrastructure.
\end{abstract}


\end{frontmatter}

\section{Motivation: Why Biomedical Knowledge Graphs?}
\label{sec:motivation}

Biomedicine has become a data-intensive science.
Large-scale initiatives such as the Cancer Genome Atlas (TCGA)~\cite{tomczak2015tcga}, the
UK Biobank~\cite{sudlow2015ukbiobank}, the Human Cell Atlas (HCA)~\cite{hca2017}, and the Clinical Proteomic Tumor
Analysis Consortium (CPTAC)~\cite{edwards2015cptacportal} have produced deep, large-scale datasets. Simultaneously, decades of curation have yielded thousands of specialized databases:
gene function repositories (Gene Ontology~\cite{geneontology2019},
UniProt~\cite{uniprot2023}), pathway databases (KEGG~\cite{kanehisa2023kegg},
Reactome~\cite{jassal2020reactome}), drug--target catalogs
(DrugBank~\cite{wishart2018drugbank}, ChEMBL~\cite{mendez2019chembl}), variant archives
(ClinVar~\cite{landrum2018clinvar}, COSMIC~\cite{tate2019cosmic}), and phenotype ontologies
(HPO~\cite{kohler2021hpo}, MONDO~\cite{mondo2022}). The biomedical knowledge base is vast but
\emph{fragmented}.

Each database was built independently, optimized for a specific domain, and operates under
its own data model, identifier namespace, evidence standard, and update cycle. A researcher
asking \emph{``which genes mutated in this patient's tumor are known drug targets, and which drugs are currently in clinical trials?''} must query at least five separate databases, resolve
synonyms, map between incompatible identifiers, and adjudicate conflicting evidence, a
process that is slow, error-prone, and not reproducible at scale.

\emph{Knowledge graphs} (KGs) address this fragmentation by representing biological
entities and their relationships in a single, traversable graph structure where:
\emph{nodes} represent biological entities (genes, proteins, drugs, diseases, pathways,
cell types); \emph{edges} represent typed, directed relationships (\textit{causes},
\textit{treats}, \textit{interacts\_with}, \textit{expressed\_in}); and \emph{properties}
on nodes and edges carry quantitative attributes and provenance metadata. Once integrated,
a KG enables multi-hop graph traversals crossing biological layers, graph-based machine
learning over heterogeneous types, and automated reasoning over curated ontological
relationships~\cite{hansel2023kgrwd}. These capabilities underpin applications from drug
repurposing and clinical decision support to mechanistic process modeling and digital twins.

However, realizing this potential depends on solving a difficult upstream problem:
\emph{data harmonization}. Before heterogeneous biomedical data can be loaded into a KG,
it must be cleaned, standardized, deduplicated, and semantically aligned.

Existing overviews of biomedical KGs concentrate on
what the resulting graphs contain and how they are used~\cite{hansel2023kgrwd}, on KG construction
methods in general~\cite{hofer2024kgconstruction}, or on curation and quality
assurance~\cite{cortes2025kgquality}. This article differs in three respects. (i)~It treats KG
\emph{assembly} as a software engineering problem and examines it through one lens: whether a KG can
be rebuilt, versioned, and updated from pinned sources. (ii)~It contrasts the practices of mature web
engineering with eight concrete engineering gaps in biomedical data integration
(Section~\ref{sec:se_challenges}), five of which are stated as explicit open problems with minimal requirements.
(iii)~It grounds the argument in a comparative reproducibility profile of six KGs
(Section~\ref{sec:examples}) and a first-hand attempt to reuse a published KG
(Section~\ref{sec:reuse}). The remainder is organized as follows: Sections~\ref{sec:harm_overview}
and~\ref{sec:harm_challenges} define the harmonization problem; Section~\ref{sec:applications}
motivates it through applications; Sections~\ref{sec:examples} and~\ref{sec:discussion} analyze
existing KGs and derive design principles; Sections~\ref{sec:gap}--\ref{sec:stack} develop the
engineering maturity gap, the eight challenges, and a target stack;
Section~\ref{sec:viz} discusses visualization; Section~\ref{sec:limitations} states limitations.

\section{The Scope of Biomedical Data Harmonization}
\label{sec:harm_overview}


\begin{figure}[h]
  \centering

\resizebox{\linewidth}{!}{%
\begin{tikzpicture}[x=0.0125cm,y=-0.0125cm]

\node[root] (root) at (672,66) {Biomedical Data\\Harmonization};

\foreach \cx in {145,489,826,1258}{
  \draw[line] (672,128) -- (672,172) -- (\cx,172) -- (\cx,211);
}

\node[cat, fill=cOrange, minimum width=3.63cm, minimum height=1.19cm] at (145,258) {Batch Effects};
\node[cat, fill=cGreen,  minimum width=3.63cm, minimum height=1.19cm] at (490,258) {Modality\\Heterogeneity};
\node[cat, fill=cPurple, minimum width=3.75cm, minimum height=1.19cm] at (837,258) {Semantic\\Heterogeneity};
\node[cat, fill=cRed,    minimum width=4.56cm, minimum height=1.19cm] at (1232,258) {Cross-Study\\Integration};

\branch{cOrange}{30}{368}{445}{525}%
  {Platform Variability}{Lab-to-Lab Differences}{Batch Correction}

\branch{cGreen}{372}{370}{455}{534}%
  {Omics vs. Imaging}{Scale \& Dimension\\Differences}{Feature Alignment}

\branch{cPurple}{715}{370}{447}{526}%
  {Different Ontologies}{Metadata Discrepancies}{Standardization}

\branch{cRed}{1078}{368}{446}{524}%
  {Cohort Variability}{Longitudinal Alignment}{Data Imputation}

\end{tikzpicture}
}

  \caption{Taxonomy of biomedical data harmonization challenges spanning four
    problem categories: \textit{batch effects}, \textit{modality heterogeneity},
    \textit{semantic heterogeneity}, and \textit{cross-study integration}. For each
    category, the first two items name typical issues and the third a common mitigation.}
  \label{fig:harmonization}
\end{figure}

Data harmonization in biomedicine operates across multiple dimensions simultaneously
(Figure~\ref{fig:harmonization}):

\begin{itemize}
    \item \emph{Batch effects} arise from technical variability in data
    generation (e.g., differences in laboratory protocols, sequencing platforms, or imaging
    scanners) that introduces systematic biases unrelated to the biological signal.
    Statistical correction methods (ComBat~\cite{johnson2007combat}, Harmony~\cite{korsunsky2019harmony}, scVI~\cite{lopez2018scvi}) address platform variability,
    but require careful design to avoid removing biological variation alongside technical
    artifacts.

    \item \emph{Modality heterogeneity} refers to the challenge of integrating
    data types, such as omics profiles, imaging signals, and clinical records, that differ in
    dimensionality, scale, and statistical properties.

    \item \emph{Semantic heterogeneity} covers inconsistencies in metadata, terminology,
    and data representation across datasets and institutions. Different ontologies and coding
    standards (ICD-10, SNOMED CT, HPO, OMIM)~\cite{haendel2018classification} prevent datasets from being directly combined.

    \item \emph{Cross-study integration} is required when datasets come from multiple
    cohorts and institutions with different patient populations, experimental designs, and
    sampling strategies.
\end{itemize}

Figure~\ref{fig:kg_pipelines} places harmonization within two related but distinct engineering workflows: \emph{KG construction and management} (left) and \emph{model assembly from knowledge graphs} (right). While both workflows involve data integration, harmonization, validation, and versioning, they differ substantially in scope, objectives, maintenance practices, and tooling requirements.

\emph{KG construction} focuses on building and maintaining a broad knowledge infrastructure intended to capture as much relevant domain knowledge as possible from heterogeneous sources such as hospitals, laboratories, publications, and repositories. The resulting graph is designed to support multiple applications. This process typically requires a major initial community effort involving ontology mapping, schema design, normalization, storage infrastructure, APIs, and user-facing services. However, long-term maintenance often becomes problematic: resources may lose active support over time, or remain curated only by closed expert groups with limited mechanisms for community feedback, collaborative editing, bug reporting, or external contributions comparable to open-source software ecosystems. As a result, the produced KG may not be optimized for the requirements of specific modeling tasks despite its broad coverage.

In contrast, \emph{model assembly} targets a specific scientific or analytical objective and therefore relies on selective extraction and integration of only the data relevant to that purpose. Instead of constructing a comprehensive graph, developers repeatedly investigate multiple KGs, datasets, and measurement repositories to curate smaller task-specific subsets. Although harmonization occurs at a smaller scale, it becomes a recurring and labor-intensive activity because each new model may require different combinations of sources, dependencies, and annotations. This workflow is characterized by rapid iteration and frequent generation of model variants. Unlike large KG initiatives, model developers often lack centralized coordination, common annotation standards, documentation guidelines, or dedicated data stewardship expertise. Consequently, many assembled models remain insufficiently documented and difficult to make fully FAIR (Findable, Accessible, Interoperable, and Reusable)~\cite{wilkinson2016fair} and CURE (Credible, Understandable, Reproducible, and Extensible)~\cite{sauro2026cure}.


\begin{figure}[ht]
  \centering

\resizebox{\linewidth}{!}{%

\begin{tikzpicture}[
  item/.style={draw=gray!60, fill=gray!6, rounded corners=2pt, align=center,minimum width=2.4cm, minimum height=1.1cm, text width=2.2cm,
               font=\footnotesize, line width=0.5pt}
]
\column{0}{kgblue}{Knowledge Graph Construction \& Management}
\sect{0.4}{-1.3}{kgblue}{Data Sources}
  {Hospitals}{Labs}{Papers}
\sect{0.4}{-3.8}{kgblue}{Collection \& Harmonization}
  {Normalization}{Ontology Mapping}{Schema Design}
\sect{0.4}{-6.3}{kgblue}{Storage \& Access}
  {Graph Database}{REST API}{User Portal}
\sect{0.4}{-8.8}{kgblue}{Validation \& Publication}
  {Version Control}{Quality Evaluation}{Documentation}
\sect{0.4}{-11.3}{kgblue}{Challenges}
  {Data Updates}{Error Correction}{Collaborative Editing}

\column{10}{kggreen}{Model Assembly from Knowledge Graphs}
\sect{10.4}{-1.3}{kggreen}{Source Selection}
  {KG1}{KG2}{Measurements}
\sect{10.4}{-3.8}{kggreen}{Data Integration}
  {Curate \& Merge}{Version Tracking}{Dependency Management}
\sect{10.4}{-6.3}{kggreen}{Model Construction}
  {Pathway Simulation}{Hypothesis Generation}{Exploratory Dataset}
\sect{10.4}{-8.8}{kggreen}{Validation \& Publication}
  {Consistency Testing}{FAIR \& CURE}{Versioned Release}
\sect{10.4}{-11.3}{kggreen}{Challenges}
  {Reproducibility}{Update Alerts}{Data\\Consistency}

\end{tikzpicture}
}
  \caption{Engineering pipelines for KG-based biomedical knowledge infrastructure.
    \textit{Left}: KG construction and management pipeline.
    \textit{Right}: Model assembly pipeline selecting and integrating data from
    multiple KGs and measurement repositories.}
  \label{fig:kg_pipelines}
\end{figure}

\section{Core Challenges in Biomedical Data Harmonization}
\label{sec:harm_challenges}

\emph{Data harmonization} encompasses five interdependent challenges:
\textit{identifier mapping}, \textit{entity resolution}, \textit{schema alignment}, \textit{evidence integration}, and \textit{provenance tracking}. Whereas Figure~\ref{fig:harmonization} classifies the sources of heterogeneity, these five challenges are the tasks a KG construction pipeline must perform to resolve them, chiefly those arising from semantic heterogeneity and cross-study integration. Table~\ref{tab:challenges} provides an overview; the following subsections describe each challenge and state-of-the-art approaches to their mitigations.

\begin{table}[ht]
\centering
\scriptsize
\caption{Five core biomedical data harmonization challenges.}
\setlength{\tabcolsep}{2pt}
\renewcommand{\arraystretch}{1.4}
\begin{tabular}{p{2.5cm} p{4.0cm} p{6.4cm}}
\toprule
\textbf{Challenge} & \textbf{Core Problem} & \textbf{Current Mitigation} \\
\midrule
Identifier Mapping    & Same entity, multiple namespaces       &
  HGNC, BioMart, Ensembl REST, Bioregistry CURIEs \\
Entity Resolution     & Same concept, no shared ID             &
  MONDO, ChEBI, EFO, embedding-based fuzzy matching \\
Schema Alignment      & Heterogeneous data models              &
  Biolink Model, RO, OWL ontologies \\
Evidence Integration  & Contradictory / multi-source claims    &
  Open Targets scoring, confidence weights \\
Provenance Tracking   & Untraceable node/edge origin           &
  PROV-O, Biolink knowledge source properties \\
\bottomrule
\end{tabular}
\label{tab:challenges}
\end{table}

\subsection{Identifier Mapping}
\label{subsec:idmap}

The most pervasive harmonization problem is \emph{identifier heterogeneity}. Every major
database assigns its own identifier to each biological entity.
The human gene \textit{TP53} and its products are referenced by at least six common identifiers, which denote \emph{related but distinct} entity types: \code{HGNC:11998} (HGNC), \code{ENSG00000141510} (Ensembl), and \code{7157} (NCBI Entrez) identify the \emph{gene}; \code{NM\_000546} (RefSeq) identifies one of its several \emph{transcripts}; \code{P04637} (UniProt) identifies its \emph{protein} product; and \code{191170} (OMIM) identifies an OMIM entry for the gene. When TCGA mutation data (Ensembl IDs), KEGG pathway data (Entrez IDs), and DrugBank drug--target data (UniProt accessions) are loaded into the same KG without resolution, \textit{TP53} may be represented by separate, unconnected nodes. Consequently, multi-database graph queries that traverse these identifiers may fail to retrieve connected information.

Identifiers of the \emph{same} entity type should be merged (here, the Entrez and Ensembl gene identifiers). Identifiers of \emph{different} types must instead be connected by typed relationships (e.g., gene--transcript and gene--encodes--protein).

Identifier spaces also evolve: Ensembl retires gene IDs on annotation updates; HGNC symbols
change; RefSeq accession version numbers increment. A KG built from data downloaded at
different times may contain phantom duplicates. Liftover tools handle genomic coordinate
versioning, but no equivalent automated tool exists for all identifier types, requiring
explicit version-pinning in KG construction pipelines.

HGNC symbols~\cite{hgnc2023} are recommended as
human-readable node labels with HGNC IDs as stable primary keys. BioMart, the Ensembl REST
API, and the UniProt ID Mapping service provide cross-reference tables. The Bioregistry
and Identifiers.org maintain canonical CURIE prefix registries
(e.g., \code{hgnc:11998})~\cite{hoyt2022bioregistry,mcmurry2017identifiers} enabling namespace-aware deduplication.
Cross-references should be stored per entity type (e.g., \code{biolink:Gene}, \code{biolink:Transcript}, \code{biolink:Protein}) and used to merge records only within a type.

\subsection{Entity Resolution}
\label{subsec:er}

\emph{Entity resolution} handles the harder case: determining whether two records refer
to the same real-world entity when no shared identifier is available. This problem is most
severe for disease and chemical entities. A single disease concept may appear as
\code{breast carcinoma} (NCIt:C4872), \code{breast cancer} (OMIM:114480),
\code{malignant neoplasm of breast} (ICD-10:C50), and \code{Breast Neoplasms}
(MeSH:D001943). Without resolution, these become four disease nodes, and queries
connecting drugs to ``breast cancer'' will miss edges from all other representations.

Entity resolution errors include \textit{false splits}, which  produce redundant nodes and incomplete subgraphs, and \textit{false merges}, which introduce factually
incorrect edges that propagate throughout queries and ML models. 
Despite the availability of ontologies, the community currently lacks a scalable collective
mechanism for propagating discovered errors back into canonical resources.
Ontologies are maintained by small curatorial teams on grant cycles; AI-assisted agents
capable of continuously inspecting entity resolution quality and proposing corrections at
scale represent a promising direction, but require community governance frameworks not yet
broadly established.

The Mondo Disease Ontology (MONDO)~\cite{mondo2022} merges terms from OMIM, Orphanet, NCIt, DOID, and MeSH into a unified hierarchy.
Ontology-grounded resolution using MONDO, EFO~\cite{malone2010efo}, and
HPO~\cite{kohler2021hpo} reduces false merges compared to
string-similarity heuristics alone. Ontology mapping services~\cite{noy2009ontology} and
LLM-assisted semantic mapping~\cite{kokash2026ontology} further assist dataset entity resolution.

\subsection{Schema Alignment}
\label{subsec:schema}

Even after identifiers are resolved, source databases model the same biological reality
using heterogeneous schemas. A drug--protein relationship may be a directed edge in
DrugBank, a TSV row in ChEMBL, or a linked-data triple elsewhere. Beyond structural
differences, sources differ in the granularity at which they model the same concept:
KEGG~\cite{kanehisa2023kegg} pathway annotations operate at the gene level;
Reactome~\cite{jassal2020reactome} at the protein isoform level; STRING~\cite{szklarczyk2021string}
interaction scores aggregate multiple evidence types into a single weight. Granularity
mismatches that are not explicitly modeled produce semantic violations invisible to
query engines.

The Biolink Model~\cite{unni2022biolink} is the most
widely adopted standard for biomedical KGs, defining 60+ node types and a controlled
predicate vocabulary. In practice, many large KGs define bespoke extensions that diverge
from any single standard, and no schema has achieved universal adoption.
AI-assisted schema mapping agents~\cite{parciak2024schema,santos2025harmonization}
represent an emerging path to reducing manual overhead.

\subsection{Evidence Integration and Conflict Resolution}
\label{subsec:evidence}

Biomedical knowledge is probabilistic and evolves. ClinVar may classify a variant as ``Pathogenic'' while a newer study reclassifies it as ``Likely Benign.'' \emph{Evidence integration} formalizes the combination of multiple evidence streams into confidence-weighted edges. The Open Targets Platform~\cite{ochoa2021opentargets} exemplifies
this, computing association scores for gene--disease pairs by integrating genetic, somatic, molecular, pathway, and text-mining evidence.

A KG built by ingesting sources at different times will contain temporally inconsistent information; without versioned provenance, stale edges silently corrupt downstream analyses.

\subsection{Provenance Tracking}
\label{subsec:provenance}

\emph{Provenance} records the full lineage of every node and edge: source database,
version used, download timestamp, transformation pipeline, and supporting evidence.
Provenance is a functional requirement for reproducibility, auditability, update management,
and confidence weighting. 

The W3C PROV-O ontology~\cite{moreau2013provo} and the Biolink
\code{primary\_knowledge\_source} property pattern~\cite{unni2022biolink} provide standard
frameworks. In practice, many published biomedical KGs record provenance only partially (e.g., Hetionet and PrimeKG; Table~\ref{tab:kg_profiles}) due to engineering complexity, which limits their utility in clinical and regulatory settings.

\section{Applications of Biomedical Knowledge Graphs}
\label{sec:applications}

Biomedical KGs enable practical applications. Understanding the intended application is essential for correct KG design: which node types to include, which evidence
thresholds to apply, how to model uncertainty, and what query patterns to optimize.

\emph{Drug Discovery and Target Identification.} The pipeline from target identification to approved drug is long and costly: a recent estimate puts the median R\&D investment per newly approved medicine at roughly \$1~billion (mean about \$1.3~billion)~\cite{wouters2020drugcosts}, and only about 14\% of drug development programs entering phase~I ultimately reach approval~\cite{wong2019clinicalsuccess}. KGs enable
multi-evidence target scoring integrating genetic, functional genomic, molecular, and clinical
evidence into ranked candidate targets~\cite{ochoa2021opentargets}. Drug
repurposing~\cite{pushpakom2019repurposing} identifies new indications for approved drugs by
finding graph paths connecting a known drug, via shared mechanism, to a new disease.

\emph{Precision Oncology and Patient Stratification.} KGs support patient stratification by
aggregating somatic mutation data~\cite{tate2019cosmic,tomczak2015tcga}, drug sensitivity
profiles, and clinical outcome data into a unified structure. Tools such as
OncoKB~\cite{chakravarty2017oncokb} serve as knowledge backbones for Molecular Tumor Board
platforms.

\emph{Biomedical Digital Twins.} A biomedical digital twin is a computational model of an
individual patient updated with clinical measurements and used to simulate disease progression
\textit{in silico}~\cite{venkatesh2022digitaltwins,laubenbacher2022digitaltwins}. KGs serve
as the biological knowledge substrate: patient-specific omics measurements are overlaid onto
the KG's relational structure, and mechanistic simulation engines use KG topology to define
model structure.

\emph{Mechanistic Process Modeling.} KG-derived pathway structures serve as the basis for
mathematical models of biological processes: ODE systems, Boolean network models of cell-fate
decisions~\cite{karlebach2008boolean}, and flux balance analysis of metabolic networks~\cite{orth2010fba}.
Integrating context-specific omics data to condition pathway topology on cell-type context is
an active research area.

\emph{Clinical Decision Support and Translational Research.} KG-powered clinical tools can
flag sequenced tumor variants as clinically actionable, suggest targeted therapy, and retrieve
relevant clinical trial identifiers. The EMBL-EBI Open Targets Platform~\cite{ochoa2021opentargets}
and the NCATS Translator~\cite{fecho2025translator} are public platforms that support such translational use.

\emph{Epidemiology and Population Genomics.} KGs integrate Genome-Wide Association Study (GWAS) results~\cite{visscher2017gwas,buniello2019gwas} with Electronic Health Record (EHR) data and environmental exposure
measurements to model complex diseases. KG infrastructure enables automated Phenome-Wide
Association Studies (PheWAS) by providing a pre-integrated graph of variant--phenotype
association edges.

\section{Representative Biomedical Knowledge Graphs}
\label{sec:examples}

This section profiles six representative KGs to illuminate design trade-offs in their construction pipelines. The key question is: how was each assembled, and can that process be reproduced and adapted? 

Table~\ref{tab:kg_profiles} summarizes the six systems across four dimensions. A KG was included if it (i) is described in a peer-reviewed publication and is publicly accessible; (ii) integrates biomedical knowledge from multiple sources; (iii) documents, at least in part, how it is built (code, scripts, or a build description), so that reproducibility can be judged from public evidence; and (iv) together with the other selected systems covers contrasting construction strategies. Each system was assessed on the four dimensions of the table using public documentation and code repositories, and reproducibility was classified qualitatively as a fully published pipeline, a pipeline re-executed from scratch at each release, reconstructible from public sources, or a curated export without an automated rebuild path. 

\begin{table*}[ht]
\centering
\scriptsize
\caption{Six representative biomedical knowledge graphs: scale, primary identifier
namespaces, pipeline reproducibility, and key limitations.}
\setlength{\tabcolsep}{2pt}
\renewcommand{\arraystretch}{1.5}
\begin{tabular}{L{1.8cm} L{3.4cm} L{2.4cm} L{3.4cm} L{3cm}}
\toprule
\textbf{KG (version)} &
\textbf{Purpose and scale} &
\textbf{Identifier namespaces} &
\textbf{Pipeline reproducibility} &
\textbf{Key limitations} \\
\midrule

Hetionet~\cite{himmelstein2017hetionet} (v1.0, Feb.\ 2017) &
Drug repurposing; $\approx$47K nodes (11 types), 2.25M edges (24 types); Neo4j + TSV/JSON &
HGNC (genes), DrugBank (compounds), DOID (diseases) &
Source-specific Python ETL code for its 29 public resources is publicly versioned; pipeline can be re-run against pinned source versions &
Static snapshot; no update mechanism; edge provenance limited to source and license properties; disease taxonomy predates MONDO \\

NCATS Translator KG2~\cite{wood2022rtxkg2} (RTX-KG2 v2.10.3, Feb.\ 2026) &
Translational research reasoning; $>$6M nodes, $>$100M edges from $\approx$70 databases; served via ARAX/TRAPI &
Biolink-compliant CURIEs throughout; cross-source merging via SRI Node Normalizer &
Fully automated, publicly available build pipeline (JSON-Lines ETL modules); designed for periodic re-runs with updated sources &
Full-graph ML training computationally expensive; quality varies across 70+ automated sources \\

Open Targets~\cite{ochoa2021opentargets} (Platform 26.06, Jun.\ 2026) &
Evidence-scored gene--disease--drug associations for drug discovery; $\approx$61K genes, 23K diseases, 16K drugs &
Ensembl (genes), EFO (diseases), ChEMBL/DrugBank (drugs) &
Open-source Apache Spark pipeline regenerates data each release; Parquet files and GraphQL API publicly downloadable &
Schema specialized to target--disease--drug triangle; non-human biology excluded \\

Monarch Initiative~\cite{putman2024monarch} (Nov.\ 2023) &
Cross-species phenotype--genotype integration for disease gene discovery; millions of edges from $>$30 sources &
MONDO (diseases), HPO/MPO/ZP (phenotypes), Biolink predicates &
KGX interchange format; cross-species similarity edges computed by OBO ontological reasoning; build process publicly documented &
Variable completeness of model-organism phenotype annotations; cross-species mappings are approximate \\

Reactome~\cite{jassal2020reactome} (v97, Jun.\ 2026) &
Manually curated biochemical pathways; $\approx$14K proteins in 2,600+ pathways, $\approx$13K reactions &
UniProt (proteins), ChEBI (small molecules); publication-level provenance per reaction &
Curated Neo4j + RDF/OWL export; graph topology (PhysicalEntity$\to$Reaction $\to$Pathway) directly usable as mechanistic model wiring &
Coverage limited to reactions with published experimental evidence; slower update cycle than automated databases \\

PrimeKG~\cite{chandak2023primekg} (v2.1, May 2022) &
ML-ready precision medicine KG; $\approx$129K nodes, 4.05M edges (30 relation types); designed for GNN training &
MONDO (diseases), DrugBank (drugs), NCBI Entrez (genes) &
Flat CSV edge lists + node feature tables; reproducible from public sources; includes train/val/test splits and pre-computed node features &
Static 2022 snapshot; limited provenance metadata; not validated for clinical deployment \\
\bottomrule
\end{tabular}
\par\smallskip\noindent\parbox{0.97\linewidth}{\scriptsize\textit{Note.} Sizes and characteristics are those reported in the cited publication or in the documentation of the release described.}
\label{tab:kg_profiles}
\end{table*}
Hetionet and RTX-KG2 publish full Extract-Transform-Load (ETL) pipelines under open licenses. Open Targets goes further with a versioned build system re-executed from scratch each release cycle. PrimeKG occupies a middle position: distributed as static CSV files, it is nonetheless reconstructible because every source is publicly accessible. Reactome publishes curated exports without an automated reconstruction path, reflecting the irreducible cost of expert curation. Monarch shows how this position can shift over time: the 2016 release already published its ETL code (Dipper)~\cite{mungall2017monarch}, but the build ran through a multi-stage RDF and SciGraph pipeline; the current Monarch KG replaces it with a simplified, modular, and automated build (monarch-ingest/Koza) that is re-executed for each monthly release~\cite{putman2024monarch}.

\section{Design Principles for Harmonization-First KG Construction}
\label{sec:discussion}

Several cross-cutting principles emerge for engineering teams building new biomedical KGs:

\begin{enumerate}
    \item \emph{Choose canonical identifier namespaces before writing any ETL code.}
    Retrospective identifier normalization is expensive and error-prone. Establish the
    primary key namespace for each node type before ingesting any source.

    \item \emph{Use a standardized data model as your schema contract.} A stable,
    community-maintained schema provides a shared vocabulary enabling interoperability
    with minimal integration effort. The Biolink Model is one
    such approach.

    \item \emph{Treat provenance as a first-class data requirement.} Every element must
    carry \code{source} and \code{source\_version} properties. Provenance is the foundation of reproducibility and auditability.

    \item \emph{Design for update, not for snapshot.} Build ETL pipelines that are
    idempotent and source-version-aware so that incremental updates can be re-ingested
    without full graph reconstruction.

    \item \emph{Prioritize the assembly pipeline over the deployed artifact.} A
    versioned, documented build pipeline that others can run to assemble a comparable KG
    from source provides far more scientific value than a static deployed instance that
    cannot be reproduced, customized, or updated.
\end{enumerate}

\subsection{Scale and Federated Access}
\label{sec:scale}

A prominent trend in large-scale KG development is toward \emph{virtualization and federated access}, driven by the practical difficulty of centralizing the volumes of data involved. The scale of recent centralized graphs illustrates this difficulty: iKraph processes over 34M PubMed abstracts across 40 public databases,
yielding more than 10M unique entities and 30M relations~\cite{zhang2025ikraph}, and the Louisiana Tumor Registry KG prototype comprises about 91 million
RDF triples ($\sim$16\,GB)~\cite{hasan2019registrykg,hasan2020registrykg}. In our assessment, strict schema enforcement is likely to become computationally costly at this scale, and \textit{best-effort semantics} may replace strict correctness.

Federated querying across heterogeneous KGs is a key composition mechanism. QLever~\cite{bast2017qlever}
provides a high-performance SPARQL engine for very large RDF datasets; the dblp KG
demonstrates cross-graph composition via QLever-powered federated SERVICE
clauses~\cite{ackermann2024dblp}. Among centralized efforts, the Data Distillery KG harmonizes over 180 ontologies
into a unified, queryable environment~\cite{ahooyi2025datadistillery}, and the Petagraph
framework, whose schema the Data Distillery KG reuses, comprises over 32 million nodes and
118 million relationships~\cite{stear2024petagraph}.

\subsection{Case Study: Reusing the Data Distillery KG}
\label{sec:reuse}

Large-scale harmonization efforts often suffer from reproducibility limitations.
As a concrete illustration from our own experience attempting to reuse the Data Distillery
KG, we encountered three recurring obstacles:

\begin{itemize}
    \item \emph{Redeployment}: The KG is distributed as a Docker container running a
    fixed-version Neo4j image. Academic HPC centers often prohibit standard Docker
    deployments due to root-privilege requirements. We had to extract KG files from the container manually, a non-trivial process due to format changes between Neo4j versions.

    \item \emph{Poor discoverability}: The schema consisted of a single node label
    \texttt{Concept} with around 1600 distinct relationship types, with data provenance
    encoded as a node property. Absent indexes made many queries run excessively long
    or fail to complete.

    \item \emph{Incomplete coverage}: In attempting to reproduce multi-hop queries
    connecting disease phenotypes to compounds via gene regulation, we found that the
    harmonization effort covers only portions of the underlying datasets. Moreover, the dozens of
    integration scripts are not consolidated into a reusable build
    pipeline~\cite{cfde_datadistillery_github}.
\end{itemize}

The KG was obtained from the project repository~\cite{cfde_datadistillery_github} (accessed 2026-05-21) in the distribution described in~\cite{ahooyi2025datadistillery}. We used the 29~August~2025 release, the most recent one at the time of access; the repository also contains a 3~January~2025 release, and a further release (9~July~2026) appeared after our experiments. The Docker image ran a Neo4j~5.x release, which was not the latest at the time of access (Neo4j~2026.x). We attempted to (a)~deploy it on the academic HPC cluster without root privileges, (b)~inspect its schema, and (c)~reproduce a series of multi-hop queries linking disease phenotypes to compounds through gene regulation. These obstacles are not unique to Data Distillery; similar friction has been reported across large-scale biomedical KG projects~\cite{hofer2024kgconstruction,cortes2025kgquality, wratten2021reproducible}, suggesting they reflect structural properties of the problem.

\subsection{Federated Ontology-Based Data Access}
\label{sec:obda}

One architectural response to these problems is the \emph{federated, ontology-based data access (OBDA)} paradigm, which creates a unified query layer over disparate sources without physical data centralization.
Approaches to KG composition draw on components of three different kinds: (i)~\emph{semantic data models}, which define the entity types and predicates used to describe knowledge but neither store data nor execute queries; (ii)~\emph{KG construction frameworks and pipelines}, such as BioCypher~\cite{lobentanzer2023biocypher} and the build processes behind Monarch~\cite{mungall2017monarch,shefchek2020monarch} and RTX-KG2~\cite{wood2022rtxkg2}, which assemble source data into a (usually centralized) graph conforming to such a model; and (iii)~\emph{query federation systems}, such as SPARQL \texttt{SERVICE} federation over engines like QLever and the TRAPI-based NCATS Translator architecture, which route queries across independently hosted resources. 

Federated ontology-based data access combines (i) and (iii). Federated KGs commonly rely on heterogeneous harmonization mechanisms with distinct reproducibility profiles: manually curated mappings, statistical batch-correction methods that are deterministic under fixed inputs but sensitive to parameters and reference cohorts, bespoke transformation scripts that are rarely reusable, and increasingly, LLM- and AI-agent-based approaches for entity resolution and schema mapping~\cite{santos2025harmonization,parciak2024schema}.

The latter introduces additional methodological concerns. LLM outputs are stochastic, model weights evolve on vendor-controlled schedules, inference behavior changes across versions, and the provenance of harmonization decisions is difficult to audit or reproduce independently. Consequently, the harmonization layer of large biomedical KGs is a continually changing, partially opaque process, with direct implications for reproducibility and long-term maintainability.

\section{The Engineering Maturity Gap}
\label{sec:gap}

A software engineer joining a new web project in 2026 will typically encounter a highly standardized development workflow. Dependency specifications are commonly maintained in files such as \code{package.json} or \code{pyproject.toml}, enabling automated installation of versioned software packages. Continuous integration pipelines routinely execute test suites, while documentation can often be generated directly from source code annotations and typed interfaces.

In contrast, constructing a KG requires the integration of data from multiple independently maintained biomedical resources. Relevant datasets may include protein structure predictions from AlphaFold, bioactivity measurements from ChEMBL, gene dependency data from DepMap, somatic mutation profiles from TCGA, and pathway information from Reactome. These resources are distributed across separate platforms and frequently differ in access methods, file formats, identifier systems, metadata conventions, and release schedules (Table~\ref{tab:sources} lists representative examples). As a result, data integration typically relies on substantial manual effort to harmonize identifiers, reconcile schema differences, track provenance, and accommodate changes introduced by upstream data providers. Unlike software package ecosystems, there is generally no unified mechanism for dependency management, semantic versioning, or automated detection of schema-level changes across the collection of resources used in a KG construction pipeline.

Web development accumulated its infrastructure over decades of incremental investment by practitioners
sharing common problems on public platforms. Biomedical data engineering is a younger and
more fragmented discipline in which large consortia tend to build bespoke pipelines rarely
abstracted into reusable tooling~\cite{wratten2021reproducible}. The consequence is that
significant scientific value is locked behind brittle and largely irreproducible data
engineering work~\cite{djaffardjy2023developing}. 

\begin{table}[ht]
\centering
\scriptsize
\caption{Representative data sources for a targeted cancer KG, illustrating format and
access heterogeneity.}
\setlength{\tabcolsep}{2pt}
\renewcommand{\arraystretch}{1.4}
\begin{tabular}{L{2.8cm} L{4.7cm} L{2.4cm} L{2.0cm} L{0.8cm}}
\toprule
\textbf{Data Source} & \textbf{Description} & \textbf{Formats} &
  \textbf{Access} & \textbf{Size} \\
\midrule
Metabolomics WB         & Metabolite concentrations across studies  &
  mwTab, .csv, .json    & Download / API & 1 GB \\
DepMap / Achilles 25Q3 & CRISPR essentiality, CNV, expression     &
  .csv, .hdf5           & Download / API & 10 GB \\
Human Protein Atlas v25 & Protein/RNA expression by tissue and cell &
  .tsv.zip, .csv        & Download   & 5 GB \\
STRING v12.0 / BioGRID 5.0.254 & Protein--protein interactions            &
  .tsv, .mitab, .xml    & Download / API & 30 GB \\
ChEMBL 36               & Bioactive molecules, assay activities    &
  .sdf.gz, .tsv, MySQL  & FTP / API  & 35 GB \\
GTEx v10                & eQTL data, RNA-seq matrices per tissue   &
  .txt.gz, .gct, .vcf   & Download / API & 50 GB \\
AlphaFold DB v6 & Predicted 3D protein structures          &
  .pdb, .cif, .tar      & FTP / Cloud & 25 TB \\
TCGA / GDC & Tumor multi-omics: MAF, RNA-seq, methylation &
  .tsv, .maf, .bam      & Gated API & 2.5 PB \\
\bottomrule
\end{tabular}
\par\smallskip\noindent\parbox{0.97\linewidth}{\scriptsize\textit{Note.} Metabolomics Workbench has no numbered releases. Data were retrieved in February 2026.}
\label{tab:sources}
\end{table}

\section{SE Challenges in Biomedical Knowledge Infrastructure}
\label{sec:se_challenges}

Software engineering practices have a critical yet under-appreciated role in biomedical research; below we list challenges which, if addressed, could make building and consuming biomedical KGs a less daunting engineering task.

\sechallenge{No Package Manager for Biomedical Data Sources}
\label{ch:pkg}

In a typical web engineering project, a Node.js developer writes \code{npm install react@18.2.0} and receives a versioned, dependency-resolved, integrity-hash-verified module tree in seconds. A bioinformatician ingesting TCGA RNA-seq data must write custom download scripts, manually track which GDC data release was used, and rely on informal documentation to
detect schema changes.

Modern package ecosystems (npm, pip, Cargo) provide four capabilities largely absent from the biomedical data landscape: \textit{semantic versioning contracts} (MAJOR.MINOR.PATCH with well-defined compatibility guarantees); \textit{dependency resolution}; \textit{integrity verification} (cryptographic hashes); and \textit{centralized registries}.

Biomedical databases are available in the form of \term{releases} which may carry a release number and human-readable release notes, but rarely include semantic versioning, machine-readable changelogs, formal contracts specifying which schema parts were updated, and mechanisms to alert consumers when a field has been renamed or deprecated.

The Bioregistry~\cite{hoyt2022bioregistry} maintains
a controlled vocabulary of identifier namespaces. The NCATS Translator's SRI Node Normalizer~\cite{fecho2025translator}
provides cross-namespace entity resolution for a subset of entity types. The BioCypher
framework~\cite{lobentanzer2023biocypher} moves toward structured adapter-based KG
construction, but stops short of full version-locked dependency management.

\begin{gapbox}{Open problem: biomedical data package manager}
A complete biomedical data package manager requires: (1) a canonical versioning scheme
for database releases; (2) a manifest format for declaring data dependencies with version
constraints; (3) a registry of published data packages with integrity hashes; and
(4) a resolver that satisfies dependency declarations while enforcing identifier namespace
compatibility.
\end{gapbox}

\sechallenge{Identifier Namespaces Without Type Enforcement}
\label{ch:ns}

In a typical Java project, \code{com.google.common.collect.ImmutableList} is unambiguous regardless of what other libraries define a class called \code{ImmutableList}: the fully qualified name is the contract. In biomedical KGs, \code{TP53} may simultaneously appear as an HGNC gene symbol, an NCBI Gene label, a UniProt gene name, and a COSMIC cancer gene reference within the same graph, often without fully reliable automatic namespace disambiguation.

The Biolink Model addresses identifier heterogeneity via CURIEs whose prefix identifies the namespace. The SRI Node Normalizer maps equivalent CURIEs to a single canonical identifier. For human genes, proteins, and a growing subset of diseases, CURIE-based resolution is functional. For metabolites, imaging features, cell lines, and numerous other resources, coverage is uneven.
In software, a reference that cannot be resolved stops the build with a linker error; data integration has no such safeguard. A join on identifiers from mismatched namespaces raises no error: it simply matches fewer rows, or none.
The pipeline fails silently, producing a KG that looks syntactically correct 
but is semantically corrupt.

\begin{gapbox}{Open problem: namespace type safety}
A system in which namespace types are first-class objects in the pipeline language, so
that joining \code{ensembl:ENSG*} with \code{ncbigene:*} identifiers produces a
compile-time type error. The OMOP CDM~\cite{overhage2012omop} enforces coding system constraints at the schema level, but no general-purpose biomedical pipeline
framework enforces namespace type safety end-to-end.
\end{gapbox}

\sechallenge{Format Heterogeneity and the Absent Canonical Intermediate Representation}
\label{ch:format}

A targeted cancer KG assembly project may need to integrate a dozen structurally incompatible formats, including FASTA, VCF, MAF, mzML, HDF5, \texttt{.mcool}, mwTab, PSI-MI XML, OBO, and plain TSV~\cite{ison2013edam}. Data sources come with different assumptions regarding coordinate systems, measurement units, reference genome versions, metadata conventions, and identifier namespaces. This situation resembles pre-standardization phases in other engineering disciplines, although biomedical data formats additionally encode domain-specific semantics that resist purely syntactic normalization.

A \term{canonical intermediate representation} (CIR), i.e., a unified internal schema or exchange model into which ingestion adapters transform heterogeneous inputs, is standard in compiler engineering (LLVM IR) and ETL systems. Candidate solutions exist at multiple layers of the stack: the Biolink Model at the semantic schema layer; RDF/Turtle and JSON-LD~\cite{sporny2020jsonld} at the representation layer; and SHACL~\cite{knublauch2017shacl} at the validation layer. The KGX format~\cite{kgx} provides a TSV-based serialization for Biolink-typed nodes and edges, while BioCypher~\cite{lobentanzer2023biocypher} formalizes source adapters as reusable typed components.

\begin{gapbox}{Open problem: validated graph interchange specification}
A specification combining: (1) a Biolink-typed node and edge schema; (2) a SHACL constraint layer expressing biological integrity rules; (3) a PROV-O provenance subgraph; and (4) an identifier namespace declaration per node type. This could enable automated KG assembly pipeline validation and compliance checking in CI/CD.
\end{gapbox}

\sechallenge{Model Format Proliferation}
\label{ch:dsl}

Biomedical mechanistic models are described in a proliferation of domain-specific languages designed for expressiveness within their domain but not for interoperability with each other or with KGs (Table~\ref{tab:bio_models}). A researcher assembling a digital twin of a cancer cell needs pathway topology from Reactome (Neo4j/RDF), kinetic
parameters from BioModels (SBML), protein--protein interaction evidence from STRING (TSV/MITAB), and gene expression context from GTEx (GCT matrices); these resources use incompatible data models, identifier namespaces, and computational frameworks. The COMBINE archive~\cite{bergmann2014combine} coordinates across SBML/CellML/SED-ML/BioPAX, and community efforts to harmonize semantic annotations across model formats~\cite{neal2019harmonizing} improve their interoperability, but no universal modeling language spanning relational KG and dynamic model behavior
has achieved consensus.

\begin{table}[ht]
\centering
\scriptsize
\caption{Examples of biomedical modeling and exchange formats and their domain focus.}
\setlength{\tabcolsep}{2pt}
\renewcommand{\arraystretch}{1.4}
\begin{tabular}{L{2.6cm} L{9cm}}
\toprule
\textbf{Model / DSL} & \textbf{Description} \\
\midrule
SBML & Widely used standard for ODE and reaction-network models. \\
BioPAX & OWL-based pathway knowledge representation. \\
PSI-MI XML / MITAB & HUPO Proteomics Standards Initiative formats for protein interaction evidence. \\
CellML & XML-based mathematical models of cell biology. \\
BNGL / BioNetGen & Rule-based language for combinatorially complex signaling networks. \\
\bottomrule
\end{tabular}
\label{tab:bio_models}
\end{table}

\sechallenge{Service Composition Beyond REST APIs}
\label{ch:services}

Biomedical data access is fragmented across REST APIs, FTP archives, SPARQL endpoints, Cypher-queryable Neo4j instances, GraphQL APIs, and gated secure enclaves. The NCATS Translator's \emph{TRAPI protocol} provides a standardized JSON message format for multi-hop biomedical queries~\cite{fecho2025translator}; Listing~\ref{lst:trapi_example}
illustrates a two-hop TRAPI query and an abridged response.
The response is schematic: the resource identifier (\texttt{infores:example-kp}) and the score are illustrative. It shows the three parts of a complete two-hop response: knowledge-graph nodes with Biolink categories and edges for \emph{both} hops (\texttt{e\_result\_0}, \texttt{e\_result\_1}); the provenance of each edge (\texttt{sources}); and a result that binds the query-graph identifiers (\texttt{n0}--\texttt{n2}, \texttt{e0}, \texttt{e1}) to knowledge-graph elements. The Open Targets GraphQL API demonstrates that introspective, typed schemas enable programmatic service discovery at KG scale~\cite{ochoa2021opentargets}.

\lstdefinelanguage{json}{
    basicstyle=\ttfamily\small,
    stringstyle=\color{blue},
    commentstyle=\color{gray},
    morestring=[b]",
    showstringspaces=false,
    frame=single,
    backgroundcolor=\color{gray!5}
}

\begin{lstlisting}[
  language=json,
  basicstyle=\scriptsize\ttfamily,
  breaklines=true,
  frame=single,
  caption={Two-hop TRAPI query and a schematic response: small molecules affecting genes associated with type 2 diabetes.},
  label={lst:trapi_example}
]
{
  "message": {
    "query_graph": {
      "nodes": {
        "n0": { "ids": ["MONDO:0005148"], "categories": ["biolink:Disease"] },
        "n1": { "categories": ["biolink:Gene"] },
        "n2": { "categories": ["biolink:SmallMolecule"] }
      },
      "edges": {
        "e0": { "subject": "n1", "object": "n0",
                "predicates": ["biolink:gene_associated_with_condition"] },
        "e1": { "subject": "n2", "object": "n1", "predicates": ["biolink:affects"] }
      }
    }
  }
}
{
  "message": {
    "knowledge_graph": {
      "nodes": {
        "MONDO:0005148":    { "name": "type 2 diabetes mellitus", 
                              "categories": ["biolink:Disease"] },
        "NCBIGene:5468":    { "name": "PPARG", "categories": ["biolink:Gene"] },
        "DRUGBANK:DB01132": { "name": "Pioglitazone", 
                              "categories": ["biolink:SmallMolecule"] }
      },
      "edges": {
        "e_result_0": {
          "subject": "NCBIGene:5468", "object": "MONDO:0005148",
          "predicate": "biolink:gene_associated_with_condition",
          "attributes": [{ "attribute_type_id": "biolink:score", "value": 0.94 }],
          "sources": [{ "resource_id": "infores:example-kp", 
                        "resource_role": "primary_knowledge_source" }]
        },
        "e_result_1": {
          "subject": "DRUGBANK:DB01132", "object": "NCBIGene:5468",
          "predicate": "biolink:affects",
          "sources": [{ "resource_id": "infores:example-kp",
                        "resource_role": "primary_knowledge_source" }]
        }
      }
    },
    "results": [{
      "node_bindings": {
        "n0": [{ "id": "MONDO:0005148" }],
        "n1": [{ "id": "NCBIGene:5468" }],
        "n2": [{ "id": "DRUGBANK:DB01132" }] },
      "analyses": [{ "resource_id": "infores:example-kp",
        "edge_bindings": { "e0": [{ "id": "e_result_0" }],
                           "e1": [{ "id": "e_result_1" }] } }]
    }]
  }
}
\end{lstlisting}

\sechallenge{Reproducible Pipelines and the Data Versioning Gap}
\label{ch:workflows}

The primary engineering factors affecting the reproducibility of biomedical knowledge graphs (KGs) are: (1) implicit software environment dependencies not captured in pipeline specifications; (2) mutable input data, including database releases updated without corresponding version changes; (3) non-deterministic tools; and (4) incomplete provenance records.

Despite the proliferation of academic and industrial workflow management systems, their adoption in KG assembly pipelines remains limited. Nextflow~\cite{ditommaso2017nextflow} and Snakemake~\cite{molder2021snakemake} have made
significant progress on causes (1) and (4), providing workflow execution models,
containerized task environments, and provenance graphs. Nextflow's dataflow model enables
natural parallelization across many samples, and its integration with containers and HPC
environments aligns well with reproducible biomedical
workflows~\cite{djaffardjy2023developing,ahmed2021design}. Listing~\ref{lst:nextflow}
illustrates a Nextflow process normalizing GTEx RNA-seq matrices in parallel.

\lstdefinestyle{nextflow}{
  basicstyle=\ttfamily\small,
  keywordstyle=\color{blue}\bfseries,
  commentstyle=\color{gray}\itshape,
  stringstyle=\color{red!60!black},
  numbers=left,
  numberstyle=\ttfamily\tiny\color{gray},
  numbersep=10pt,
  frame=single,
  framerule=0.4pt,
  rulecolor=\color{gray!40},
  backgroundcolor=\color{gray!5},
  breaklines=true,
  tabsize=2,
  xleftmargin=16pt,
}

\begin{lstlisting}[style=nextflow,  label={lst:nextflow},
  caption={Nextflow process normalizing GTEx RNA-seq matrices in parallel.}]
params.gtex_release = 'v10'  // pinned data release

process NORMALISE {
  tag "GTEx ${params.gtex_release}: ${gct.baseName}"
  container 'python:3.11.9-slim'   // pinned image
  input:
    tuple path(gct), val(sha256)
  output:
    path "${gct.baseName}.norm"
  script:
  """
  echo "${sha256}  $gct" | sha256sum -c -
  python normalise_matrix.py "$gct" > "${gct.baseName}.norm"
  """
}
\end{lstlisting}

Listing~\ref{lst:nextflow} pins the data release and the container image and verifies each input matrix against a SHA-256 checksum (the normalization script is a placeholder). Even so, this verification is hand-written in each process, and the workflow engine has no notion of a versioned data dependency: a bare declaration such as \texttt{path gct} does not verify that the GTEx matrix is the version the process was designed for. The analogous infrastructure for \textit{data sources}, namely versioned, hash-verified packages with well-defined access interfaces, does not yet exist in general-purpose form for biomedical data. 

\begin{gapbox}{Open problem: data dependency management for workflows}
A layer for workflow systems that: (1) declares upstream data sources with pinned version
identifiers in the workflow manifest; (2) verifies integrity of downloaded data via published checksums; (3) logs data source versions into the provenance record alongside
compute environment hashes; and (4) alerts when data is updated by checking
registered release feeds.
\end{gapbox}

\sechallenge{KG Lifecycle Management: Schema Evolution and Sustainability}
\label{ch:lifecycle}

Biomedical databases suffer a cluster of lifecycle failures well understood in software
engineering but not systematically addressed in the biomedical context:

\begin{itemize}
    \item \emph{Silent schema evolution.} As noted for SE Challenge~1, database releases may rename or remove columns without providing a machine-readable changelog. Downstream pipelines that hard-code field names may consequently return missing or misaligned data without raising errors, with the problem detected only months later.

    \item \emph{Identifier retirement without forwarding.} Gene identifiers, protein accessions, and compound IDs are periodically retired or merged without automated forwarding mechanisms (the biological equivalent of HTTP 301 redirects), leaving dangling references in KG nodes.

    \item \emph{Funding discontinuity.} Multiple scientifically important databases have experienced funding gaps, resulting in API outages or stale data. Public biomedical data services generally lack uptime SLAs and standardized graceful deprecation policies.
\end{itemize}

BioCypher~\cite{lobentanzer2023biocypher} represents the most structured current approach,
making schema changes visible as code diffs in versioned adapter code. The broader
sustainability problem is structural: unlike software packages maintainable by a single
developer at near-zero marginal distribution cost, biomedical KGs require ongoing
consortium-scale curation investment.

\sechallenge{Scaling from Experimental Pipeline to Production}
\label{ch:devtoprod}

The transition from an experimental pipeline to full-cohort production deployment is one
of the most consistently underestimated engineering challenges in biomedical data
integration and a documented source of lost development
time~\cite{djaffardjy2023developing,wratten2021reproducible}.

\begin{itemize}

    \item \emph{Scale changes the nature of the problem.} Scaling biomedical KG construction introduces challenges related to data quality, identifier consistency, and execution concurrency that may remain undetected in smaller workflows. Pipeline steps well-studied for one-shot execution have received little systematic attention in the context of incremental updates~\cite{hofer2024kgconstruction}. Furthermore, knowledge graphs pursuing similar goals exhibit substantial variation in source integration and terminology~\cite{cortes2025kgquality}. Integration of EHR data introduces further complexity, as standardized graph representations for clinical data sources remain underdeveloped~\cite{hansel2023kgrwd}. Multi-hop retrieval over large biomedical KGs has itself become a major computational bottleneck~\cite{cheng2026kgscaling}.

    \item \emph{Testing infrastructure is immature or absent.} Standard software engineering prescribes three layers of testing: \emph{unit tests} for individual processing steps; \emph{integration tests} verifying that chained steps produce correct outputs; and \emph{end-to-end tests} on representative data slices that can run in CI/CD. Two partial solutions are \texttt{nf-test}~\cite{forer2025nftest} and \texttt{NFTest}~\cite{patel2024nftest}, which address unit and snapshot testing of Nextflow processes. Test-driven development of KG pipelines remains uncommon compared to general software engineering projects.

    \item \emph{Failure observability.} When a full-cohort pipeline fails after consuming hundreds of CPU-hours, developers typically receive only an exit code, with little information about the specific input record or pipeline state at the time of failure.
\end{itemize}

\begin{gapbox}{Open problem: production-readiness engineering for biomedical pipelines}
Concretely absent are: (1) a \emph{staging environment specification}, i.e., a miniaturized
but structurally faithful replica enabling full pipeline runs in under two hours; (2) a \emph{structured failure logging contract}, i.e., a standard schema for recording failure context to a queryable store; (3) a \emph{within-task checkpoint API} for
partial recovery; (4) a \emph{data contract validation layer} that performs automated schema checks at each stage boundary. The nf-core community~\cite{ewels2020nfcore} has begun
institutionalizing some of these practices, but a coherent framework applicable across workflow systems and biomedical data modalities is still missing.
\end{gapbox}

\section{Toward a Unified Engineering Stack for Biomedical Knowledge}
\label{sec:stack}

Most KG preparation efforts prioritize domain-specific challenges, treating engineering aspects as secondary concerns addressed through ad hoc, project-specific solutions. Consequently, engineering practices remain fragmented and often incompatible across projects. The community lacks a coherent, general-purpose engineering stack that integrates and streamlines the full lifecycle of KG construction, maintenance, and model assembly (Table~\ref{tab:stack}).

\begin{table}[ht]
\centering
\scriptsize
\caption{The engineering maturity gap in biomedical knowledge infrastructure across eight
challenge areas.}
\setlength{\tabcolsep}{2pt}
\renewcommand{\arraystretch}{1.4}
\begin{tabular}{L{3cm} L{3.4cm} L{7cm}}
\toprule
\textbf{SE Concept} & \textbf{Partial Biomedical Equivalent} & \textbf{What Is Missing} \\
\midrule
Package manager & Bioregistry, BioCypher adapters &
  Semantic versioning, dependency resolution, integrity hashes for data releases \\
Namespace / type system & Biolink CURIEs, SRI Node Normalizer &
  Namespace type enforcement at pipeline compile time \\
Canonical IR & KGX format, Biolink Model &
  Full SHACL constraint layer; biological integrity rules; provenance subgraph spec \\
Universal modeling language & SBML / BioPAX / COMBINE standards &
  Unified language spanning relational KG and dynamic model behavior \\
Typed service contract & TRAPI protocol, Open Targets GraphQL &
  Semantic input/output types; automated composition from ontological type signatures \\
CI/CD + immutable envs & Nextflow / Snakemake + containers &
  Data source version pinning and integrity verification in workflow manifests \\
Schema migration & BioCypher adapter versioning &
  Machine-readable schema changelogs; automated downstream breakage detection \\
Staging / prod parity & nf-core CI/CD standards, nf-test &
  Staging environment spec; structured failure logging; within-task checkpoint API \\
\bottomrule
\end{tabular}
\label{tab:stack}
\end{table}

The convergence trajectory is emerging: the Biolink Model and CURIE prefix registries provide the schema and namespace contract;
the NCATS Translator ecosystem provides a composition protocol; BioCypher and KGX provide
structured ETL abstractions; Nextflow with containers provides reproducible compute environments; JSON-LD
and SHACL provide web-native semantic data exchange. What is missing is their
\emph{integration into a coherent, documented stack}, together with an
engineering culture that ships pipelines alongside snapshots. The goal is a
workflow that lets a researcher reproduce or customize a KG in a few commands from
version-pinned sources, the way a developer runs \code{npm install} to reconstitute a
web application.

\section{Visualization as a First-Class Concern}
\label{sec:viz}

One aspect of the biomedical data engineering stack that receives comparatively little
investment is \emph{visualization as infrastructure}. Most KG platforms provide only rudimentary reactive graph views, typically
force-directed layouts where nodes and edges are distinguished primarily through color and
shape. The highly informative diagrams found in scientific publications, e.g., illustrating
pathways, anatomical hierarchies, or multi-scale biological organization, are mostly created manually and cannot be reproduced or automatically updated when the underlying
data change.

The visualization literature provides many diagram types suited to structured knowledge
representation: trees, treemaps, directed acyclic graphs, heatmaps, and visualizations
combining graphs with quantitative plots. 
Modern visualization frameworks such as D3.js~\cite{bostock2011d3},
Vega-Lite~\cite{satyanarayan2017vegalite}, Observable Plot~\cite{observableplot},
and Apache ECharts~\cite{li2018echarts} make it possible to generate such diagrams
programmatically~\cite{mei2018design}, but the required data transformation
pipelines are rarely standardized or integrated into KG infrastructures.

An emerging opportunity lies in LLM-assisted visualization: a language model could
interpret a domain scientist's natural language query, decompose it into a structured
retrieval and transformation plan, and dispatch sub-tasks to specialized microservices
via the \emph{Model Context Protocol (MCP)}~\cite{anthropic2024mcp}, a standardized
interface enabling AI agents to invoke external tools in a composable, auditable manner.
Realizing this vision requires visualization pipelines to be registered as FAIR resources
with explicit semantic descriptions of their input--output contracts.

A \emph{semantic visualization layer}, in which node and relation types are mapped to
visual encodings through declarative rules derived from domain ontologies, would make visualizations reproducible, queryable, and automatically updatable.
Approaches such as constraint-based layouts and structural-signature anomaly detection
illustrate how visualization could evolve into a systematic tool for knowledge validation
and quality assurance.

\begin{figure}[h]
\centering
\begin{subfigure}[b]{0.38\textwidth}
    \centering
    \includegraphics[width=\textwidth]{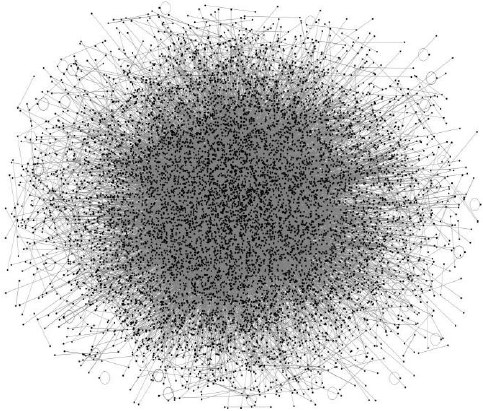}
    \caption{KG without visualization effort.}
    \label{fig:graph-vis-a}
\end{subfigure}
\hfill
\begin{subfigure}[b]{0.6\textwidth}
    \centering
    \includegraphics[width=\textwidth]{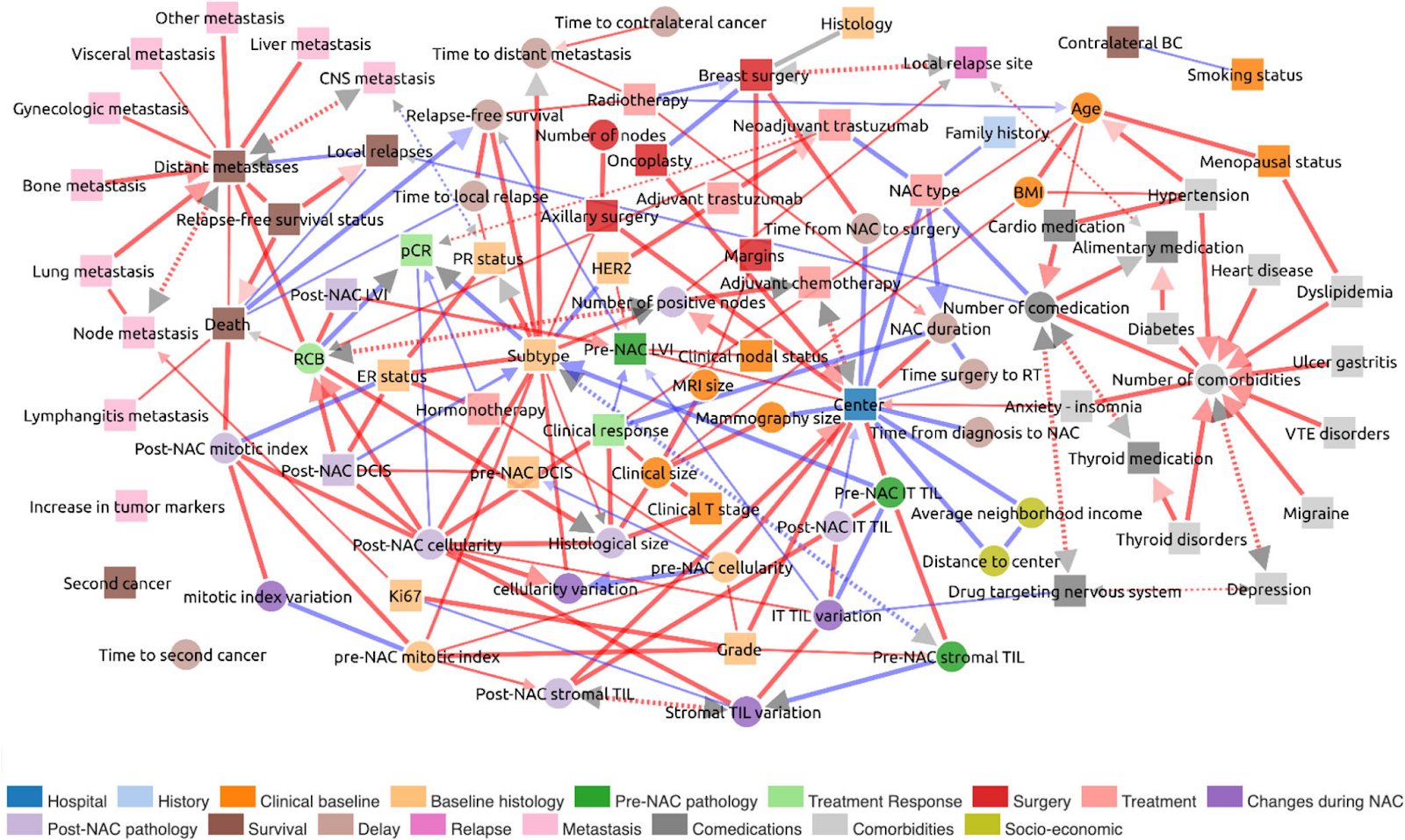}
    \caption{KG visualization with node and edge classification.}
    \label{fig:graph-vis-b}
\end{subfigure}

\vspace{0.5cm}

\begin{subfigure}[b]{0.38\textwidth}
    \centering
    \includegraphics[width=\textwidth]{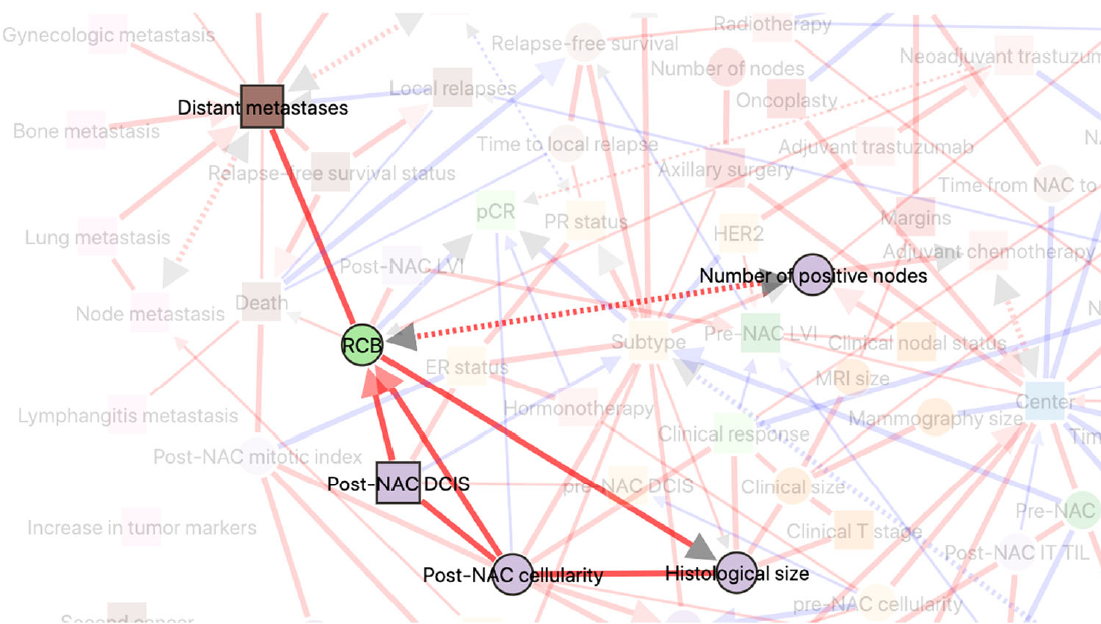}
    \caption{Reactive KG visualization.}
    \label{fig:graph-vis-c}
\end{subfigure}
\hfill
\begin{subfigure}[b]{0.6\textwidth}
    \centering
    \includegraphics[width=\textwidth]{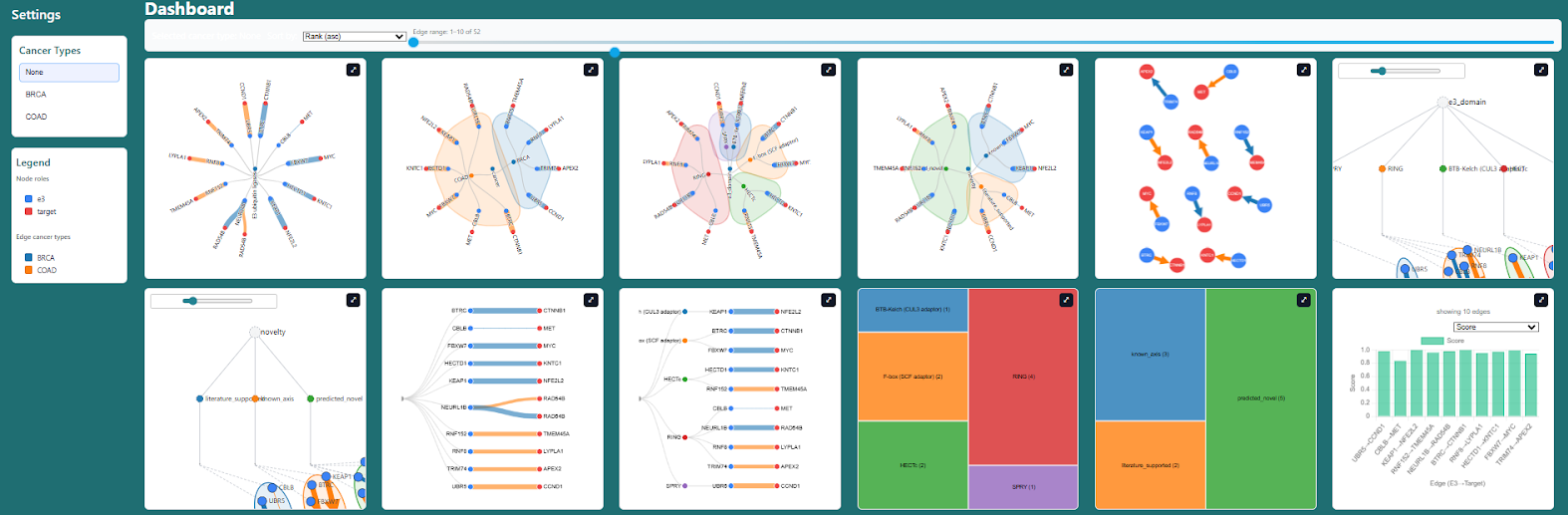}
    \caption{KG analytics via visual exploration with transformation pipelines.}
    \label{fig:graph-vis-d}
\end{subfigure}

\vspace{0.5cm}

\begin{subfigure}[b]{0.38\textwidth}
    \centering
    \includegraphics[width=\textwidth]{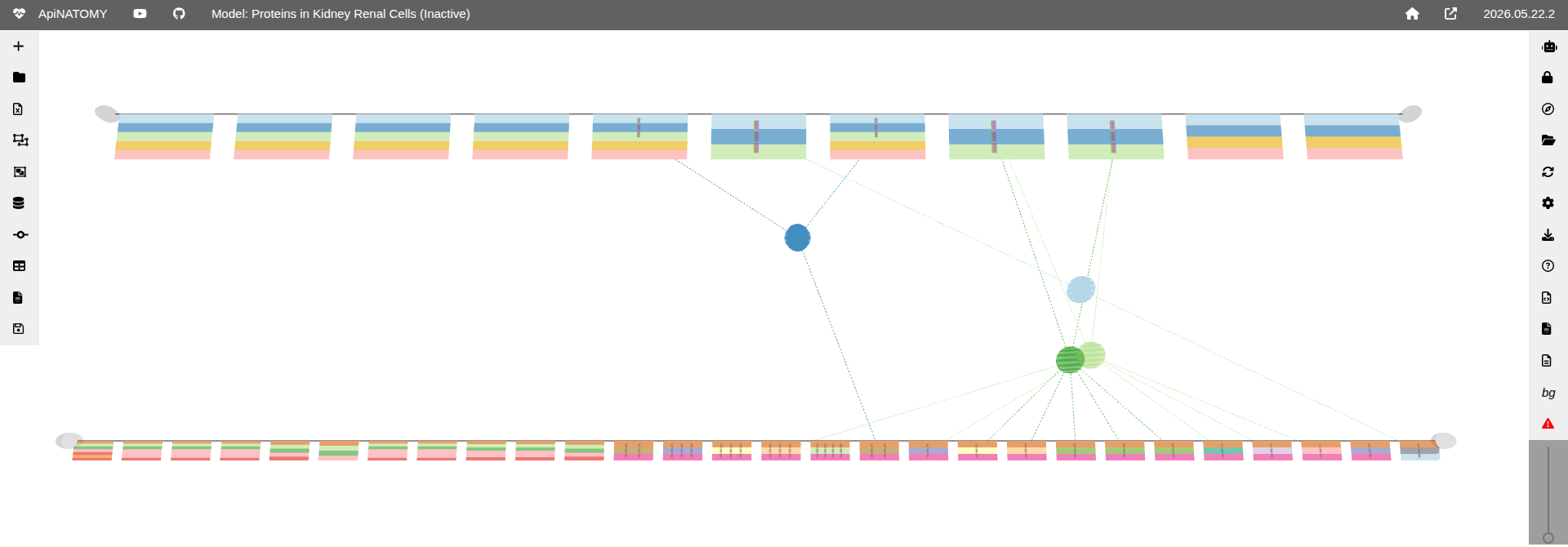}
    \caption{KG semantic- and context-aware layout generation.}
    \label{fig:graph-vis-e}
\end{subfigure}
\hfill
\begin{subfigure}[b]{0.6\textwidth}
    \centering
    \includegraphics[width=\textwidth]{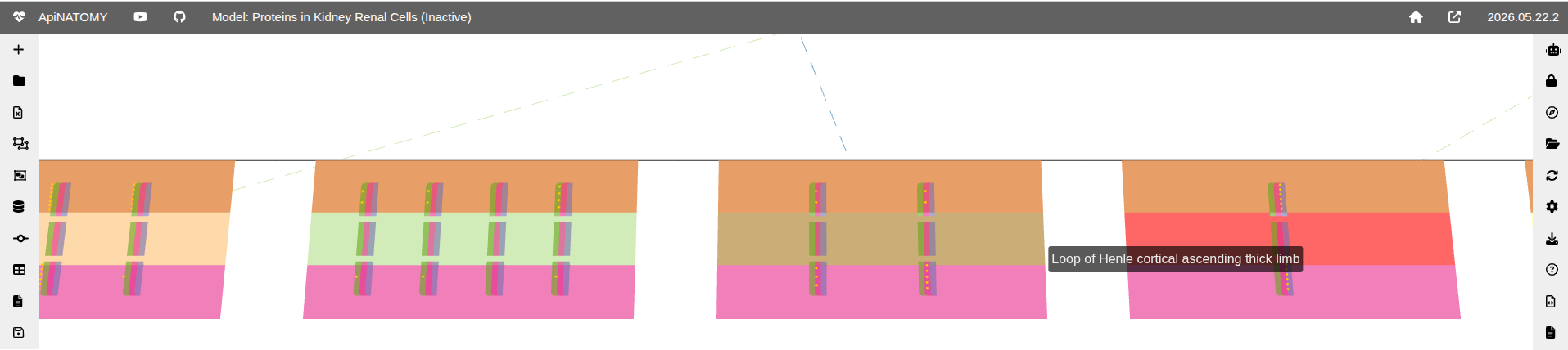}
    \caption{Relationships between concepts providing visualization constraints.}
    \label{fig:graph-vis-f}
\end{subfigure}

\caption{Levels of graph-based biomedical data visualization, from no effort (a) through
  node-edge classification (b), reactive exploration (c), transformation-based analytics (d),
  to full semantic context-aware layout (e, f).}
\label{fig:visualization}
\end{figure}

Figure~\ref{fig:visualization} illustrates how visualization effort transforms the
usability of the underlying data. The higher levels (Figure~\ref{fig:graph-vis-d}) require derivation of trees from
graphs, clustering within graph structure, and shared filters across reactive components;
such pipelines are rarely integrated with KG schemata. The ApiNATOMY panels~\cite{kokash2021knowledge} in Figures~\ref{fig:graph-vis-e} and~\ref{fig:graph-vis-f} show an approach where KG elements know how they must be drawn: ``part-of'' relationships place an entity node within boundaries of its hosting entity, and other relationship types constrain position, size, shape, and color within the force-directed layout.

In ApiNATOMY, however, most of the drawing rules are embedded in the viewer
code~\cite{kokash2025opv} and cannot be reused without reprogramming. Ideally,
ontology schemas describing a KG should be paired with declarative visualization
specifications and style sheets, so that the semantics of the data and its
graphical presentation are defined separately but distributed together, much as
HTML and CSS are on the Web.

Deagen et al.~\cite{deagen2022fair} take a step in this direction by storing Vega-Lite specifications alongside SPARQL queries in the KG, so that charts are rendered on demand from the latest data. Their specifications, however, describe
individual charts rather than presentation rules for classes and relationships, and they mainly target statistical plots. Mapping each ontology class and property to reusable layout constraints and visual encodings,  
as ApiNATOMY does for ``part-of'' relationship types, would let any diagram built from the KG inherit a consistent presentation across tools and reduce the custom code currently needed for each new diagram.

\section{Scope and Limitations}
\label{sec:limitations}

Several dimensions of biomedical knowledge infrastructure are addressed only
briefly or not at all. The article focuses on the pipeline layer, i.e., the
software engineering needed to \emph{assemble, version, and maintain} biomedical
KGs. Machine learning for KG completion, graph neural networks, clinical
prediction, biostatistics, clinical validation, and regulatory compliance are
mentioned as context but not surveyed. Likewise, the eight SE challenges in
Section~\ref{sec:se_challenges} are a curated selection that leaves out
privacy-preserving computation, federated learning governance, streaming
integration, and natural language interfaces. The six KGs in
Section~\ref{sec:examples} were chosen to illustrate design diversity, and numerous notable KGs are not profiled.

The comparison with web development is not empirical;
the article does not quantitatively evaluate the engineering quality of KG construction projects at scale. Its evidence consists of the comparative profiles, the illustrative listings, and the Data Distillery case study, which reflects one team's experience with one (fairly large) project.

\section{Conclusion}
\label{sec:conclusion}

This article has examined biomedical knowledge infrastructure from two complementary perspectives: the integration and harmonization of heterogeneous biological knowledge, and the engineering practices required to construct, maintain, and evolve biomedical KGs.

Despite their flexibility, biomedical KGs remain costly to construct and maintain, requiring substantial domain expertise and custom integration pipelines. Key challenges include identifier mapping, entity resolution, schema alignment, evidence integration, and provenance tracking. Compared with mature software development ecosystems, biomedical KG engineering lacks widely adopted mechanisms for dependency management, versioning, reproducible builds, and automated validation.

A central distinction is that between deployed KG instances and reproducible KG construction processes. While portals provide access to integrated knowledge, reproducible build pipelines enable their reconstruction, verification, customization, and extension. Future research should therefore prioritize standardized data and package management, identifier and namespace management, interoperable graph formats, machine-verifiable integrity constraints, and reproducible end-to-end integration workflows.

The emergence of LLM-based agents further increases the need for transparent, versioned, and interoperable knowledge infrastructure. Standards such as TRAPI and MCP provide interfaces for interaction with knowledge resources, while agents could assist with resource monitoring, schema alignment, and pipeline execution. However, reliable automation depends on the availability of reproducible and verifiable underlying infrastructure. Progress in biomedical knowledge engineering ultimately requires closer integration of domain knowledge, software engineering practices, and interoperable standards, supported by collaboration among biomedical researchers, knowledge engineers, software developers, and data providers.

\bibliographystyle{elsarticle-num}
\bibliography{main}

\end{document}